\documentclass[iop,apl,twocolumn,showpacs,amsmath]{revtex4}
\usepackage{graphicx}

\begin{document}

\title{First-principles study of Ti-doped sodium alanate surfaces}

\author{Jorge \'I\~niguez$^{1,2}$ and Taner Yildirim$^1$}

\affiliation{$^{1}$NIST Center for Neutron Research, National
Institute of Standards and Technology, Gaithersburg, MD 20899\\
$^{2}$Dept. of Materials Science and Engineering, University of
Maryland, College Park, MD 20742}

\begin{abstract}
We have performed first-principles calculations of thick slabs of
Ti-doped sodium alanate (NaAlH$_4$), which allows to study the system
energetics as the dopant progresses from the surface to the bulk. Our
calculations predict that Ti stays on the surface, substitutes for Na,
and attracts a large number of H atoms to its vicinity. Molecular
dynamics simulations suggest that the most likely product of the
Ti-doping is the formation of H-rich TiAl$_n$ ($n>1$) compounds on the
surface, and hint at the mechanism by which Ti enhances the reaction
kinetics of NaAlH$_4$.
\end{abstract}


\maketitle

One key issue for the advancement of fuel-cell technologies is the
development of safe and inexpensive ways of storing hydrogen aboard
vehicles. In recent years sodium alanate (NaAlH$_4$) has become one of
the most promising systems to achieve this goal. The kinetics of the
reversible reaction by which pure NaAlH$_4$ releases hydrogen is
relatively slow. However, it was discovered that a few percent of Ti
doping increases the reaction rates dramatically~\cite{bog97},
bringing the system close to what is required for practical
applications. In spite of extensive investigations, the mechanism by
which Ti enhances the reaction kinetics is still
unknown~\cite{gro02a,kiy03,gro03}. In fact, even the location of the
dopants, on the surface~\cite{gro00} or in the bulk of the
system~\cite{gro03,sun02}, and the reactions they cause remain
unclear. Some recent experimental results suggest the formation of a
TiAl$_3$ compound on the surface of the system~\cite{gra04}, but a
consensus on this issue is yet to emerge.

In this Letter we present a theoretical study addressing the following
outstanding questions. Where is Ti in doped NaAlH$_4$, on the surface
or in the bulk? If Ti stays on the surface, does it substitute for Na
or Al, and what are the reactions and structural rearrangements it
causes?


The calculations were performed within the generalized gradient
approximation~\cite{per96} to density functional theory as implemented
in the code SIESTA~\cite{sol02}. We used a localized basis set
including double-$\zeta$ and polarization orbitals, and
Troullier-Martins pseudopotentials~\cite{tro93}. We tested the
convergence of our calculations with respect to the k-point and real
space meshes. For example, for the 001-surface calculations involving
4500~\AA$^3$ supercells and about 162 atoms, we used a
2$\times$2$\times$1 k-point grid and a 150~Ry cutoff for the real
space mesh. In our SIESTA calculations some accuracy is sacrificed,
mainly because of the small basis set we use, so that bigger systems
can be studied. Hence, we checked some representative SIESTA results
with a more accurate {\sl ab initio} method, using a plane-wave basis
set and ultrasoft pseudopotentials~\cite{van90} as implemented in the
code CASTEP~\cite{pay92}.


%
\begin{figure}
\includegraphics[scale=0.45,angle=0]{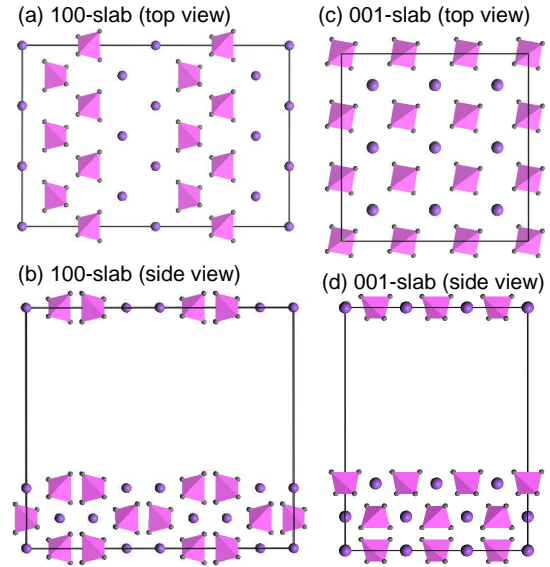}
\caption{Top and side views of the 100-slab (a-b) and the 001-slab
(c-d) supercells of NaAlH$_4$. Small and large gray spheres represent
H and Na atoms, respectively. The Al atoms are at the center of the
tetrahedra.}
\label{fig1}
\end{figure}

We started by determining which doping models of NaAlH$_4$ surfaces
are most energetically favorable. For that purpose we considered the
two slab-type supercells in Figs.~\ref{fig1}a-b and \ref{fig1}c-d,
which correspond to the most natural terminations of tetragonal
NaAlH$_4$ [along the (100) and (001) directions, respectively]. Note
that these supercells can be respectively regarded as composed of
$3\times3\times4$ and $3\times3\times3$ NaAlH$_4$ groups (i.e., 216
and 162 atoms). In both cases we imposed that the atoms in the deepest
layer be fixed at the bulk atomic positions, and allowed about 13~\AA\
of empty space between slabs. The bulk structural parameters
($a=b=5.01$~\AA\ and $c=11.12$~\AA) were obtained from
first-principles and are in good agreement with previously published
results~\cite{ini04}. We considered a structure to be relaxed when
residual force components are smaller than 0.03~eV/\AA. NaAlH$_4$ is a
fairly ionic system~\cite{agu04,pel04} and the pure surfaces do not
present particularly relevant features. The electronic energy
minimizations converge easily and the structural deviations from the
bulk are relatively small.

In order to investigate the Ti doping of these surfaces, we computed
the cohesive energy ($E_{\rm coh}$) of a number of experimentally
motivated doping models~\cite{ini04}: the substitution of Al and Na by
Ti, as well as the occurrence of Na vacancies near the dopant. We use,
for example, the notation ``Ti@Na+Na$^v$'' to refer to the case in
which Ti substitutes for Na and there is a Na vacancy next to it. In
all the calculations we considered only one Ti dopant in the
supercell.

\begin{table}
\caption{Calculated cohesive energies, given in eV and per NaAlH$_4$
formula unit, for several doping models (see text). The column marked
with an asterisk shows CASTEP results. In all the cases, $E_{\rm coh}$
of the pure surface is taken as the zero of energy. The pure $E_{\rm
coh}$ values are given in the last line.}
\vskip 5mm
\begin{tabular*}{0.90\columnwidth}{@{\extracolsep{\fill}}lrrr}
\hline\hline
Doping model       & (100)   & (001)    & (001)$^*$ \\ 
\hline
Ti @ Al            &  0.05   &  0.02    &  0.04     \\
Ti @ Al + Na$^v$   & $-$0.01 & $-$0.09  & $-$0.06   \\
Ti @ Al + 2Na$^v$  & $-$0.11 & $-$0.28  & $-$0.25   \\
Ti @ Na            &  0.11   &  0.12    &  0.13     \\
Ti @ Na + Na$^v$   &  0.04   &  0.03    &  0.04     \\
Ti @ Na + 2Na$^v$  & $-$0.01 & $-$0.07  &  $-$0.05  \\
\hline
undoped & 20.48 & 19.28 & 20.15 \\
\hline\hline
\end{tabular*}
\label{tab1}
\end{table}

The cohesive energies shown in Table~\ref{tab1} were obtained as the
sum of the individual atom energies minus the energy of the
system. For both surfaces, $E_{\rm coh}$ of the pure case is taken as
the zero of energy. Hence, the results in the table give minus the
energy change in reactions of the form Ti+Na$_{27}$Al$_{27}$H$_{108}$
$\rightarrow$ Al+Na$_{27}$Al$_{26}$TiH$_{108}$, and measure the
relative stability of pure and doped systems. A positive (and large)
entry in the table indicates that, in principle, it is feasible to
obtain the doped surface.

We find that both Ti@Na and Ti@Al are energetically more stable than
the pure surfaces, i.e., the system gains energy by accepting a Ti
dopant and releasing a Na or Al atom. In addition, Ti@Na is found to
be the most favorable substitution, and the differences between the
(100) and (001) surfaces are minor. Interestingly, the most favorable
surface doping models turn out to be the same as in the bulk
case~\cite{ini04}. We also find that the energy cost of forming Na
vacancies is relatively small, suggesting they are likely to occur in
the real doped system. Finally, Table~\ref{tab1} shows that the SIESTA
and CASTEP results are in good agreement.

The obtained structural relaxations closely resemble those described
in Ref.~\cite{ini04} for Ti-doped bulk NaAlH$_4$. In the cases where
Ti substitutes for Na, the dopant drags neighboring H atoms towards
itself and greatly stretches the corresponding Al--H bonds. The Ti for
Al substitution leads to an expansion of the TiH$_4$ group, as Ti is
larger than Al. This size difference is probably the reason that Ti@Na
has a larger cohesive energy than Ti@Al, even if the nominal valences
of the atoms would suggest otherwise.


%
\begin{figure}
\includegraphics[scale=0.4,angle=0]{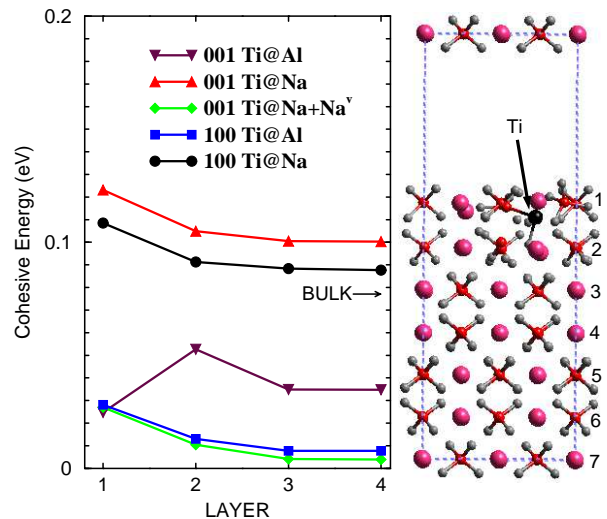}
\caption{Calculated cohesive energies of various doping models (see
text) as a function of the Ti dopant location (see right panel). In
all cases the pure surface $E_{\rm coh}$ is taken as the zero of
energy.}
\label{fig2}
\end{figure}

Next we studied whether the Ti dopant stays on the surface or, rather,
penetrates into the bulk of the system. For that purpose, we
considered slabs that are very thick along the direction normal to the
surface (see Fig.~\ref{fig2}), and computed the cohesive energy of the
system as a function of the dopant position. More precisely, we worked
with supercells formed by 7$\times$2$\times$2 and 2$\times$2$\times$7
NaAlH$_4$ groups, respectively, for the (100) and (001) surfaces. The
atoms in the deepest layer are fixed at the bulk positions. Such
supercells allow us to study the dopants up to three layers away from
the surface, which we found is enough to reach the bulk limit. We
focused on the doping models found to be more stable than the pure
surface, i.e., Ti@Na, Ti@Al, and Ti@Na+Na$^v$.

Figure~\ref{fig2} shows the results. In all cases $E_{\rm coh}$ of the
pure system is subtracted so that the value at the surface
approximately coincides with that of Table~\ref{tab1}. (The agreement
between Table~\ref{tab1} and Fig.~\ref{fig2} should not be perfect,
since the respective systems differ in dopant concentration.) Since a
larger $E_{\rm coh}$ implies greater stability, we find that in all
the cases considered it is energetically favorable for the Ti atom to
remain on the surface of the system. In fact, the dopant always finds
it favorable to locate in the outermost layer, except in the case of
Ti@Al on the (001) surface.

As mentioned above, we find pure NaAlH$_4$ surfaces to be
electronically and structurally similar to the bulk of the
material. Hence, we think the dopants stay on the surface simply
because this allows larger structural relaxations to accommodate them.


%
\begin{figure}
\includegraphics[scale=0.4,angle=0]{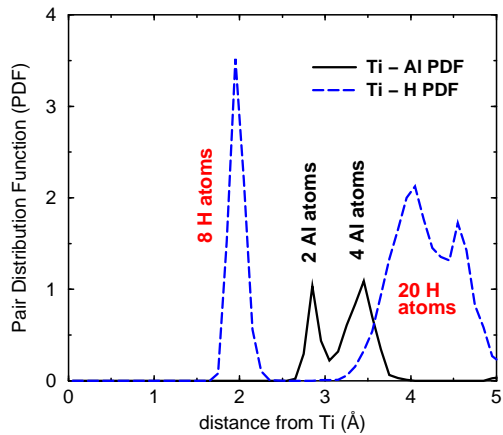}
\caption{Pair distribution function obtained from a 300~K {\sl ab
initio} MD simulation of the Ti@Na+2Na$^v$ 001-surface (see text).}
\label{fig3}
\end{figure}

Finally, in an effort to understand better the reactions and
structural relaxations that occur in real NaAlH$_4$ surfaces, we
performed {\sl ab initio} molecular dynamics (MD) simulations at
300~K. We worked with the (001) slab depicted in Fig.~\ref{fig1}c-d
and studied the doping models Ti@Na, Ti@Na+Na$^v$, and
Ti@Na+2Na$^v$. Our MD runs were about 2~ps long, which is hardly
enough to thermalize a system as reactive as this. Nevertheless, we
were able to study the average bonding of Ti by calculating pair
distribution functions (PDF's; Fig.~\ref{fig3} shows the result for
Ti@Na+2Na$^v$), and reached two distinct conclusions.

In all the cases considered Ti is surrounded by a large number of H
atoms. More precisely, the first Ti--H PDF peak is centered at about
1.9~\AA\ and corresponds to 7 hydrogens for Ti@Na and 8 for
Ti@Na+2Na$^v$, Ti@Na+Na$^v$ being an intermediate case. These
hydrogens travel continuously and freely between the dopant and the Al
atoms they were originally bound to. The presence of the dopant thus
results in a large number of loosely bound hydrogens that are
relatively close to each other. This might well facilitate H$_2$
formation and release, consequently enhancing the kinetics of the
reactions in the decomposition/formation of doped alanates. Note that
in Ref.~\cite{ini04} bulk Ti dopants were found to bind strongly a few
H atoms at 0~K, and that is probably also the case for surface
dopants. However, the dynamical situation at 300~K seems to be one in
which {\sl many} hydrogen atoms approach the dopant and none of them
is actually tightly bound.

The other effect that we observed in all our MD runs is the formation
of Ti--Al bonds. The first Ti--Al PDF peak is centered at about
2.75~\AA\ and involves 2 Al atoms in the Ti@Na and Ti@Na+2Na$^v$
cases, and almost 4 in the case of Ti@Na+Na$^v$. Taking into account
that the typical Ti--Al distance in TiAl$_3$ systems is about
2.80~\AA\ \cite{col02}, this clearly hints at the formation of some
kind of TiAl$_n$ ($n>1$) compound on the surface of the real
material. The calculations suggest such a compound should be rich in
interstitial hydrogen. Unfortunately, our supercells are not big
enough to study the size and structure of these hypothetical TiAl$_n$
compounds. Nevertheless, our results clearly support the conclusions
of the recent X-ray absorption study of Ref.~\cite{gra04}.


In summary, we have carried out a first-principles study of Ti-doped
NaAlH$_4$ surfaces. Our conclusions are: (1) It is energetically
favorable for the Ti dopants to stay on the surface rather than to
penetrate into the bulk of the material. (2) Ti substitutes
preferentially for Na, and formation of neighboring Na vacancies is
likely. (3) The Ti dopant attracts to its neighborhood a large number
of H atoms (up to 8). These hydrogens are loosely bound and close to
each other, which might facilitate H$_2$ formation and release. (4) We
find strong indications that a TiAl$_n$ ($n>1$) compound forms on the
surface of the alanate, in support of recent experimental
reports~\cite{gra04}.

We thank T.J. Udovic and C. Brown for their comments on the
manuscript.

\end{document}